\documentclass[12pt]{article}%
\usepackage{amsmath}
\usepackage{amsfonts}
\usepackage{amssymb}
\usepackage{graphicx}%
\newif{\ifcomentarios}
\comentariosfalse
\begin{document}

\title{Surfing at the wave fronts: the bidirectional movement of cargo particles
driven by molecular motors}
\author{Daniel Gomes Lichtenth\"{a}ler and Carla Goldman\thanks{Correponding author.}\\Departamento de F\'{\i}sica Geral - Instituto de F\'{\i}sica \\Universidade de S\~{a}o Paulo CP 66318\\05315-970 S\~{a}o Paulo, Brazil.}
\date{March 18, 2008}
\maketitle

\begin{abstract}
The collective behavior of molecular motor proteins have been investigated in
the literature using models to describe the long-time dynamics of a
unidimensional continuum motor distribution. Here, we consider the phenomena
related to the transport of particles (vesicles, organelles, virus, etc) in
the realm of these continuum motor systems. We argue that cargo movement may
result from its ability to perturb the existing motor distribution and to
\textit{surf} at the resulting shock waves separating regions of different
motor densities within the transient regime. \ In this case, the observed
bidirectionality of cargo movement is naturally associated with reversals of
shocks directions. Comparison of the quantitative results predicted by this
model with available data for cargo velocity allows us to suggest that
geometrical characteristics of the transported particle shall determine the
extension and intensity of the perturbation it produces and thus, its
dynamics. Possible implications of these ideas to virus movement at the cell
body are discussed in connection with their distinguished morphological characteristics.


\end{abstract}

\section{\bigskip Introduction}

The intracellular active transport of particles, including organelles,
vesicles and virus is mediated by motor proteins such as myosin, kinesins and
dyneins \textbf{\cite{howard}}. The unidirectional motion of a single motor
along protein filaments is well characterized experimentally and it was first
modeled at a microscopic level by a stochastic dynamics describing the
behavior of a Brownian particle in the presence of a time-dependent asymmetric
potential field \cite{astumian 94},\cite{astumian 96},\cite{adjari
prost},\cite{magnasco}. The idea is based on the mechanism of "ratchet and paw
" introduced by Feynmann to discuss the meaning of the second law of
thermodynamics \cite{feynman}. Since then, this model has been used as a
prototype to explain why and in what conditions Brownian particles are able to
do work against external potential gradients. Also if particles can bind
directly to motors, the transport of cargo follows as a direct consequence of
motor movement.

More recently, it appeared in the literature attempts to describe the movement
of interacting Brownian motors, since it was realized that collective effects
emerging from this situation may be relevant to explain certain
characteristics of cellular transport and, in particular, the observed
property of bidirectionallity of cargo movement \cite{gross},\cite{murray
00},\cite{welte 04},\cite{ross 06}. This non-diffusive type of process is
characterized by inversions of cargo direction after processive runs
\footnote{Processive run in this context is referred to the movement
accomplished in one definite direction before inversion or detachment from
filament.} that may be preceded by relative large \textit{resting times}
(intervals within which the particles remain at the same microtubule site) .

Attempts to describe the observed bidirectional movement of extensive objects
(filaments) mediated by motor proteins use essentially the same ideas of the
original microscopic models \cite{colective filament}. \ Numerical results
indicate that some of the observed characteristics of \ bidirectional movement
are captured in this way provided that the number of motors attached
simultaneously to the filament, arranged in a periodic way, be no less than
\ a certain critical number $N_{\min}$. Because of this, it is not clear how
such models could be extended to describe the observed bidirectional movement
of small particles, as vesicles or virus.

The most accepted explanation in these cases is referred in the literature as
the \textit{coordination model }\cite{gross} . According to this, the
bidirectional movement would result from the coordinated action of two type of
motors - a plus-ended and a minus-ended motor - attached simultaneously to
cargo (filament). The reversal of its direction would just reflect the fact
that one or other type of motor, but not both simultaneously, shall be active
during the respective time intervals. The motor coordination, that is, the
control of motors activity would be accomplished by an external non-motor
protein complex that should be able to coordinate the movement and timing of
many motors of different nature and distinct characteristics - certainly a
very non-trivial job. Such an external complex has never been identified in
any \textit{in vivo }experiment \cite{gross 02}.

At certain scales of interest, phenomena related to the collective behavior of
interacting Brownian motors can also be approached\ by models that are
intended to describe dynamic aspects of continuum media. Such description,
referred in the following as \textit{macroscopic} model, is generally
justified upon evaluation of the characteristic sizes and time scales where
the molecules operate at very low Reynold's numbers \cite{chowdhury 03}.
Usually, the approach is based on the continuum versions of the
\textquotedblleft asymmetric exclusion processes\textquotedblright\ (ASEP)
\cite{andjel},\cite{ferrari},\cite{kgrug 91},\cite{derrida} for studying the
\textit{long-time} behavior of motors interacting through volume excluded
\cite{chowdhury 03} \cite{parmeggiani 03} \cite{parmeggiani 04}
\cite{lipowski}.\textbf{ }From this perspective, the individual microscopic
asymmetric movement of motors is assumed \textit{a priori} and for
\textit{open boundary conditions} the problem consists in analyzing the
\textit{steady-state} behavior of a defined motor density as the solutions to
the corresponding non-linear differential equation - the non-viscous Burgers
equation - that describes the dynamics of the system in these limits. It is
also possible to superimpose to the ASEP a non-conservative Langmuir process
to allow the system exchange motors with the bulk at any position of the
microtubule \cite{parmeggiani 04}. Apparently, however, questions related to
cargo transport, have not been considered in this context .

Here, we make a proposal in this direction based on considerations about the
interactions between cargo and motor molecules. It relies on the idea that the
arrival of a cargo particle may perturb the motor system in such a way to
produce local changes onto the existing motor density.

Thereafter cargo may take advantage of the gradients induced by this initial
perturbation to move along microtubule by \textit{surfing} at the density
shock waves formed as the motor system relaxes back to the situation before
interaction with cargo. Shock waves separate regions of different motor
densities and evolve according to the \textit{transient} solutions to the
corresponding Burgers equation for the considered initial conditions
(perturbation). Within this view, bidirectional movement of cargo particles
follows naturally as a consequence of the reversals of shocks propagation direction.

From a more quantitative perspective, we use data from the movement of
vesicles in Drosophila embryo \cite{welte 98} to perform a phenomenological
analysis of the model considering the explicit expressions obtained for shock
velocities. As we shall see, this allows us to associate geometrical
characteristics of cargo with the extension and magnitude of the perturbation
it produces. We then discuss on the possibility to make a connection between
these results with the properties of the observed movement of virus at the
cell body.

\section{From the microscopic potential models to the macroscopic motor
density profile}

We consider the continuum limit of the mean field approximation (the
macroscopic limit) of an ASEP discrete model representing the stochastic
dynamics of particle motors moving in a one-dimensional space. Within this
view, the diffusion process is not accounted for explicitly; its effects on
the movement of each motor under the time-dependent asymmetric potential, are
assumed \textit{a priori. }Yet, the parameters that define the asymmetry of
the potential, can be incorporated into the parameters of the considered
ASEP\textit{. }

Consider the time-dependent asymmetric and periodic microscopic potential
shown in \textbf{Fig. 1}. The position of each minimum may be associated to
the position of a one-dimensional lattice site; the distance between sites
being $l=a+b$ coincides with the spacial period of the potential. The
difference between parameters $a$ and $b$ define the strength of the
asymmetry. The time dependence is such that it alternates between an "off"
state (flat potential), when the particle feels no forces, and an state \ "on"
- the sawtooth profile shown. If the potential is turned off and the particle
starts to diffuse from a local minimum corresponding to the site $i$ of the
lattice, it shall diffuse around this position up to the instant when the
potential is turned on again, after a time interval $\tau$. At this instant,
the particle is supposed to fall immediately into the nearest minimum
(adiabatic approximation). Because $a>b,$ and considering a regime of relative
slow diffusion, it is more likely that the nearest \ minimum achieved by the
particle be that at site $i+1$. This prototype mechanism was conceived to
explain from a microscopic point of view the observed unidirectional movement
of a Brownian motor \cite{astumian 96}, \cite{adjari prost}, \cite{magnasco}.

ASEP models have been introduced more recently in this context as a way to
account for collective effects of many motors moving on the same lattice model
(\cite{parmeggiani 03}, \cite{lipowski}). At these scales, motors are
\textit{self-driven} and the interactions among them are that of volume
excluded, i.e. a motor can not occupy a site already occupied by another
motor. Let us now consider a stochastic description of the process at this
scale. The time unit being the interval $\tau$ within which a particle can
execute one movement to a neighbor site with (transition) probability $p_{i}$
to move from site $i$ to site $i+1$ and (transition) probability $q_{i}$ to
move from site $i$ to site $i-1$. Neglecting correlations (mean-field
approximation) the average density $\rho_{i}(t+\tau)$ of motors at site $i$,
at time $t+\tau$ satisfies the following recurrence relation:
\begin{equation}
\rho_{i}(t+\tau)=\rho_{i}(t)+\rho_{i+1}(t)q_{i+1}(t)+\rho_{i-1}(t)p_{i-1}%
(t)-\rho_{i}(t)(p_{i}(t)+q_{i}(t)) \label{recurrence 1}%
\end{equation}
Excluded volume interaction is introduced into the model by attributing to
$p_{i}$ and $q_{i}$ an explicit dependence on the occupation of the target
sites:%
\begin{equation}
p_{i}(t)=p(1-\rho_{i+1}(t)) \label{p1}%
\end{equation}
and%
\begin{equation}
q_{i}(t)=q(1-\rho_{i-1}(t)) \label{q1}%
\end{equation}
Here, $p$ and $q$ are dimensionless constants such that $p,q<1$. Now, let
\ $N$ be the total number of lattice sites and let $L\equiv1$ be its total
length. As in Ref.\cite{parmeggiani 04} , the discrete model can be
coarse-graining with lattice constant $l=1/N$ \ to a continuum such that for
$N\rightarrow\infty$ the density $\rho_{i}(t)$ becomes a function $\rho(x,t)$
of a continuum variable $x=i/N.$ In this limit, $\rho_{i\pm1}$ can be expanded
in powers of $l:$
\begin{equation}
\rho_{i\pm1}(t)=\rho(x\pm l,t)=\rho(x,t)\pm l\partial_{x}\rho(x,t)+\frac{1}%
{2}l^{2}\partial_{x}^{2}\rho(x,t)+O(l^{3}) \label{taylor}%
\end{equation}
and the recurrence in (\ref{recurrence 1}) becomes%

\begin{equation}
\frac{\rho(x,t+\tau)-\rho(x,t)}{\tau}=\gamma\left[  \frac{(p+q)}{2}%
l\partial_{x}^{2}\rho(x,t)+(p-q)(2\rho(x,t)-1)\partial_{x}\rho(x,t)\right]
+O(l^{2}) \label{recurrence 2}%
\end{equation}
where we have defined%
\begin{equation}
\gamma=\frac{l}{\tau} \label{gama}%
\end{equation}

For $\tau\rightarrow0$ and $l\rightarrow0$ keeping $\gamma$ in (\ref{gama})
finite and neglecting terms of $O(l)$, Eq. (\ref{recurrence 2}) converges to
the non-viscous Burgers equation%

\begin{equation}
\partial_{t}\rho+\gamma(p-q)(1-2\rho)\partial_{x}\rho=0. \label{burgers 1}%
\end{equation}

Now, a connection between the above description and the microscopic model can
be made by setting
\begin{equation}
p=a/l\hspace{0.4in}\text{and }\hspace{0.4in}q=b/l \label{q e p}%
\end{equation}
so that Eq. (\ref{burgers 1}) is rewritten as%
\begin{equation}
\partial_{t}\rho+K\partial_{x}(\rho(1-\rho))=0 \label{burgers 2}%
\end{equation}
where we have defined $K\equiv\gamma(a-b)N$ \ positive. This can also be
expressed as $\partial_{t}\rho+\partial_{x}j=0$ with the particle current
given by
\begin{equation}
j(x,t)=K\rho(x,t)(1-\rho(x,t)). \label{corrente}%
\end{equation}
Notice that to keep $K$ finite as $N\rightarrow\infty$, parameters $a$ and $b$
must scale as $1/N.$ The same behavior is expected for the Langmuir rates in
the model of Ref. \ \cite{parmeggiani 04}. In these conditions, and with no
lost of generality, we can set $K=1.$

The nonequilibrium steady-state solutions to this equation have already been
explored in the references mention above to study the \textit{long - time}
behavior of molecular motor traffic with open boundary conditions. This allows
one to make important predictions on the stationary density profiles related
to the concentrations gradients of motors at the microtubule ends. In
particular, the emergence of stable domain walls separating regions of
different motor densities, predicted in these conditions, have indeed been
observed in recent experiments \cite{nishinari 05}. Here, we consider periodic
boundary and study the \textit{short-time} behavior of the system described by
Eq. (\ref{burgers 2}), after being perturbed with respect to the motor density
steady-state $\rho(x,t)=1/2$. The results we obtain for a particular choice of
initial conditions are shown in the next section.

\bigskip

\section{Results}

It is known that perturbations around the steady state solutions of Burgers
equation (\ref{burgers 2}) may create shocks due to encounters of particles at
different velocities (\cite{haberman},\cite{evans}). This leads us to
speculate that cargo may use these shock waves, by \textit{surfing }at the
wave fronts, in order to reach target sites.

To illustrate these ideas, we can generally conceive that the approach of
cargo induces a perturbation on motor density such as to create at the cargo
neighborhood regions where motor density becomes lower and others where
density becomes higher than the steady state $\rho=1/2.$ As an example, we
choose
\begin{equation}%
\begin{array}
[c]{ccc}%
\rho(x,0)=1/2 & \text{for} & x<0\\
\rho(x,0)=1/2-\varepsilon & \text{for} & 0<x<2a\\
\rho(x,0)=1/2\text{ } & \text{for} & 2a<x<3a\ \\
\rho(x,0)=1/2+\varepsilon & \text{for} & 3a<x<4a\\
\rho(x,0)=1/2 & \text{for} & x>4a
\end{array}
\label{condinic}%
\end{equation}
to represent such perturbation (see \textbf{Fig(2a)}). Parameters $a$ and
$\varepsilon$ are related, respectively, to the extension and magnitude of
this perturbation.

We then use the method of characteristics to find the transient solutions to
Eq. (\ref{burgers 2}) that reveal the time evolution of shocks for these
initial conditions (IC) (\ref{condinic}), which is our main interest here.
This is performed in the Appendix.

The results for the density profile are represented in\textbf{ Fig.(2)} at
various instants of time. The behavior of each shock front, namely, distance
traveled up to an eventual encounter with other shock fronts - which are
identified here with the processive runs - \ the time spent within each of
these runs, and the corresponding expressions for average shock velocities are
compiled in \textbf{Table1}.

In the next Section, we use these results to examine the consequences of the
hypothesis made above.

\section{\bigskip Discussion}

\subsection{the surfing}

The model proposed here offers rather simple explanations to a few but
essential phenomena related to motor-driven cargo transport. In particular, it
suggests an alternative to interpret data showing that cargo executes a
non-diffusive bidirectional movement before it arrives at its final destination.

In general the main difficulty to explain the bidirectional movement of cargo
relies precisely on the fact that the existing models have been conceived to
explain the unidirectional movement of motors. Because of this, the original
idea that the cargo movement shall be put into effect by its attachment to a
single or a few identical motor proteins has been reviewed in the literature.

The alternative mechanism proposed here does not exclude the possibility that
eventually, the presence of two types of motors may contribute to explain the
reversals of particle direction. However, it does not use this as a necessary
condition as does the coordination model mentioned above. Within such an
alternative context, cargo is driven by \textit{shocks} and this is
independent of the nature of motors involved provided they offer an assessable
density background. That is, in order to move and eventually change direction,
cargo must be able to perturb this background and follow gradients of motor
density right after perturbation, just like the putative\ microscopic "second
class particles" introduced in the context of many-particle asymmetric
exclusion processes (ASEP) to study the nature of shocks at these scales
\cite{andjel}, \cite{ferrari}.

Although we do not offer here any direct evidence that would support these
ideas we can test for their compatibility by confronting some experimental
data with the results shown in Table 1 for the considered initial conditions.
For example, it is shown there that a cargo particle accompanying the wave
front referred here as wave front 2, or second shock, would perform a backward
movement within a time interval $T_{1}=$\textbf{ }$a/2\varepsilon$ that
depends on the model parameters. Within the subsequent time interval $T_{2}=$
$a/4\varepsilon,$ the particle would remain at rest up to the instant
$3a/4\varepsilon$ when it reverses the direction of movement. Thereafter, the
particle would proceed eventually changing the magnitude of its velocity
though not direction anymore (in this specific case (\ref{condinic}) the
reversal happens just once) until the motor density recovers its steady state
profile $\rho(x,t)=1/2$ and the particle reaches its final destination at the
one of the microtubule ends, as expected. Therefore, at least qualitatively,
the phenomenological aspects of the observed movement, as reversals of
direction, processivity in both directions and resting times are accounted for
by this model.

Evidently, if one considers other initial conditions, for example by
introducing additional parameters to characterize the perturbation, it is
expected a great variety of time intervals and corresponding distances
traveled in both directions, differing in processivity and velocity.
Therefore, a relevant aspect of this proposal is the possibility to attribute
primarily to cargo the task for creating such gradients, thus inducing its own
\textit{surfing }movement. This is discussed below in connection to a more
quantitative analysis of available data from real biological systems.

\subsection{phenomenological analysis}

Consider for instance the data in the literature on the movement of lipid
vesicles on Drosophila embryo \cite{welte 98}. We can check for their
compatibility with the results in Table 1 all expressed in terms of just two
parameters $a$ and $\varepsilon$. To accomplish this, we observe that data
analysis obtained from the transport of vesicles in the referred experiment
identifies two typical runs in both directions: one that is short (average
distance $d_{short}\sim100nm$) and slow (average velocity $v_{slow}$
$\sim200~nm/s$) and another that is long (average distance $d_{long}%
\sim1000nm$) and fast (average velocity $v_{fast}\sim400~nm/s$) . We shall use
these data to\ estimate the values of $a$ and $\varepsilon.$ We proceed by
identifying $v_{slow},$ taken from experimental data, \ with the average
velocity of the 4$^{th}$ shock front $v_{4}=$ $2$ $\varepsilon/7$ . It
results\footnote{It must be remembered that the parameter $\varepsilon$ was
defined adimensional and that the velocities are expressed in units of the
parameter $K$ that have been defined equal to one.} $\varepsilon\sim700$. We
can now use this to predict the magnitude of the average shock front velocity
for the $5^{th}$ shock $v_{5}=\varepsilon/2\sim350$ which approaches very well
the values (absolute values) observed for $v_{fast}$ $\sim400nm/s.$

On the other hand, these results must be compatible with the predictions for
the distances traveled by the corresponding shocks. This complements the
analysis by estimating possible values for parameter $a.$ Accordingly, the
predicted distance $d_{5}$ must be approximately one order of magnitude
greater than $d_{4}$. This is in fact accomplished by the results shown in
Table 1 since from there $d_{5}/d_{4}=8.$

We shall now perform a slight different analysis of the data by redefining the
constant $a$ such that%

\begin{equation}
\ a\equiv500p(nm) \label{a}%
\end{equation}
where $p$ is a numerical factor. From this, $d_{4}=80p(nm)$ and $d_{5}%
=650p(nm).$ Therefore, if we choose $p=1.2$, it results $d_{4}\sim100(nm)$ and
$d_{5}\sim800(nm).$ These can be compared to the results for $d_{short}$ and
$d_{long}$ given above to conclude that the characteristic times and distances
traveled by shock fronts 4 and 5 can be related in a very direct way with the
typical values obtained in experiments for short and long runs, respectively.

More interesting, however, is the fact that this approach to the experimental
data is made possible with the choice in (\ref{a}) for $p\sim O(1)$. First,
notice that $500nm$ is the average diameter of the vesicles transported in the
referred experiments. Thus, if the nature of interactions between cargo and
motors are of short range, this implies that the magnitude and extension of
the perturbation should be related to the geometrical characteristics
\ (typical sizes) of the cargo. Therefore, if the nature of interaction of
different cargos with motors is the same, but cargo differ from each other in
morphological aspects, these differences would be reflected in the specific
way (size, magnitude) each of them perturb the equilibrium motor distribution
and consequently, in the diversity of movements. Such diversity is indeed
observed in experiments using different motor/cargo systems and/or under
different conditions \cite{gross}.

The above considerations seem specially attractive to explain the behavior of
virus particles moving within the cellular environment observed more recently
using confocal microscopy technique (\cite{lehmann}, \cite{ploubidou},
\cite{kerstin}). According to the authors of these experiments, virus appear
\textit{surfing} along filopodia in such a way to perform a bidirectional
motor-driven process in the presence of a \textit{single type of motor}
protein, unidirectional myosin II. From these observations the authors then
discuss on the possibility that, in these experiments, virus might have been
able to create the conditions to induce their own movement. These observations
support the idea that virus use (hijack) existing cell transport machinery to
invade the cell \cite{kerstin}. The model presented here suggests a mechanism
to explain how this can be accomplished without external and specific
mechanisms of coordination. Such explanation has a focus on the morphological
aspects of cargo - which, of course, is a distinguished characteristic of
virus. In our view, this complements the suggestions made in Ref.
\cite{lehmann} relative to the existence of a common mechanism that drives
different virus particles to their targets in cells, in spite of their
morphological differences. We propose that this common mechanism is just
related to their ability to interact with the system by perturbing the
existing motor profile. Specific geometries, however, would be used by virus
to create a variety of movements, possible necessary for their efficiency. We
thus believe that these ideas may be helpful for studying the mechanisms of
virus infection and also for designing drugs that must be directed to the cell body.

\bigskip

\section{Appendix}

Here we obtain the properties of shock waves shown in \textbf{Table1} within
the transient regime, after the motor system is perturbed by interaction with
cargo. We apply the method of characteristics to find the transient solutions
to the equation (\ref{burgers 2}). For this, we follow the texts by Habermann
(\cite{haberman}) and that by Evans (\cite{evans}).
\begin{equation}
\frac{\partial\rho(x,t)}{\partial t}+(1-2\rho(x,t))\frac{\partial\rho
(x,t)}{\partial x}=0
\end{equation}
for the chosen initial conditions (IC) (\ref{condinic}). In turn, these
solutions define the properties of shocks depicted in Table 1 which is our
main concern in this work. For each pair ($\rho(x_{0}),x_{0}$) there
corresponds a characteristic curve for (\ref{burgers 2}) given by
\begin{equation}
x(t)=(1-2\rho(x_{0}))t+x_{0} \label{eq carac}%
\end{equation}
Thus, the IC in (\ref{condinic}) define the different characteristic
families:
\begin{align}
x_{1}(t)  &  =x_{0}\text{ \hspace{0.15in}for \hspace{0.15in}}x_{0}%
<0\label{carac2}\\
x_{2}(t)  &  =x_{0}+2\varepsilon t\text{ \hspace{0.15in}for \hspace{0.15in}
}0<x_{0}<2a\nonumber\\
x_{3}(t)  &  =x_{0}\text{ }\hspace{0.15in}\text{for }\hspace{0.15in}%
2a<x_{0}<3a\nonumber\\
x_{4}(t)  &  =x_{0}-2\varepsilon t\hspace{0.15in}\text{ for}\hspace
{0.15in}\text{ }3a<x_{0}<4a\nonumber\\
x_{5}(t)  &  =x_{0}\hspace{0.15in}\text{ for }\hspace{0.15in}4a<x_{0}\nonumber
\end{align}
These are represented in Fig.(3) where one observes the occurrence of shocks
dividing regions of different density values. There are six shocks. There are
also shown in this figure the rarefaction regions that are localized between
any two characteristic families that get apart from each other as the time
evolves. The exact determination of the origin of each of these shocks (time
and space) - and rarefaction regions\ and the course of their time evolution
allows us to describe the evolution of $\rho(x,t).\ $

At the initial time ($t=0$) there are two shocks:

\begin{itemize}
\item \textbf{shock 1 ($x_{s1}(t)$)} at position $x_{s1}(0)=2a$, between the
families defined by $x_{2}(t)$and $x_{3}(t)$. The velocity of this shock front
is determined by mass conservation conditions :
\begin{equation}
\frac{dx_{s1}}{dt}=\frac{j(\rho(x_{3}))-j(\rho(x_{2}))}{\rho(x_{3})-\rho
(x_{2})}=\varepsilon\nonumber
\end{equation}
where the current $j(\rho)$ is defined by (\ref{corrente}). We then have
\begin{equation}
x_{s1}(t)=2a+\varepsilon t
\end{equation}
that gives the shock position as a function of time.

\item \textbf{shock 2 ($x_{s2}(t)$)} at $x_{s2}(0)=3a$, between the families
defined by $x_{3}(t)$ and $x_{4}(t)$. The velocity of this shock is given by
\begin{equation}
\frac{dx_{s2}}{dt}=\frac{j(\rho(x_{4}))-j(\rho(x_{3}))}{\rho(x_{4})-\rho
(x_{3})}=-\varepsilon\nonumber
\end{equation}
from which
\begin{equation}
x_{s2}(t)=3a-\varepsilon t
\end{equation}

\end{itemize}

Still at $t=0$, there originate two rarefaction regions

\begin{itemize}
\item \textbf{rarefaction 1 ($x_{r1}(t)$)} at $x_{r1}(0)=0$, evolving in the
region between families $x_{1}(t)$ and $x_{2}(t)$. The requirement of entropic
solutions drive us to look for the corresponding family of characteristic
curves $x_{r1}(t)$ defined by the density $\rho_{r1}$ at each point in this
region:
\begin{equation}
x_{r1}(t)=(1-2\rho_{r1})t\hspace{0.15in}\text{for }\hspace{0.15in}%
1/2-\varepsilon<\rho_{r1}<1/2
\end{equation}
which are also represented in Fig. 2. Conversely, one can write the solutions
to $\rho_{r1}(x_{r1},t)$ for density profile at each position $x_{r1}$ in this
region as \textbf{ }
\begin{equation}
\rho_{r1}(x_{r1},t)=1/2-\frac{x_{r1}}{2t}\hspace{0.15in}\text{for }%
\hspace{0.15in}0<x_{r1}<2\varepsilon t \label{raref1}%
\end{equation}

\item \textbf{rarefaction 2 ($x_{r2}(t)$)} at $x_{r2}(0)=4a$, evolving in the
region between families $x_{4}(t)$ and $x_{5}(t)$. The characteristics in this
region are given by
\begin{equation}
x_{r2}(t)=4a+(1-2\rho_{r2})t\hspace{0.15in}\text{ for }\hspace{0.15in}%
1/2<\rho_{r2}<1/2+\varepsilon
\end{equation}
thus
\begin{equation}
\rho_{r2}(t,x_{r2})=1/2+\frac{4a-x_{r2}}{2t}\hspace{0.15in}\text{ for }%
\hspace{0.15in}4a-2\varepsilon t<x_{r2}<4a \label{raref2}%
\end{equation}

\item \textbf{shock 3 ($x_{s3}(t)$)} The third shock occurs at time $t_{3}%
^{0}$ when the two families $x_{s1}(t)$ and $x_{s2}(t)$ cross. Thus,
$t_{3}^{0}$ is determined by the condition:$x_{s1}(t_{3}^{0})=x_{s2}(t_{3}%
^{0})$, which gives%
\begin{equation}
t_{3}^{0}=a/2\varepsilon.
\end{equation}

The velocity of the shock is given by
\begin{equation}
\frac{dx_{s3}}{dt}=\frac{j(\rho(x_{4}))-j(\rho(x_{2}))}{\rho(x_{4})-\rho
(x_{2})}=0\nonumber
\end{equation}
meaning that the shock front is stationary at $x_{s3}(t_{3}^{0})=x_{s1}%
(t_{3}^{0})=5a/2$.

\item \textbf{shock 4 ($x_{s4}(t)$)} occurs at time $t_{4}^{0}$ when the
family $x_{4}$ with origin at $x_{0}=4a$ reaches the second region of
rarefaction and also cross the characteristics from family $x_{2}.$ Thus,
$t_{4}^{0}$ is determined by the relation $x_{s3}(t_{4}^{0})=x_{r2}%
(\rho=1/2+\varepsilon;t_{4}^{0})$, which gives
\begin{equation}
t_{4}^{0}=3a/4\varepsilon
\end{equation}
The (continuity) condition, in this case reads $\dfrac{dx_{s4}}{dt}%
=\dfrac{j(\rho_{r2}(x_{s4}))-j(\rho(x_{4}))}{\rho_{r2}(x_{s4})-\rho(x_{4})}$ ,
that is%
\begin{equation}
\frac{dx_{s4}}{dt}=\varepsilon-\frac{4a-x_{s4}}{2t}%
\end{equation}
then $x_{s4}$ is given by the solutions to the above differential equation for
initial condition $x_{s4}(t_{4}^{0})$ such that $x_{s3}(t_{4}^{0}%
)=x_{s4}(t_{4}^{0}).$ The result is \cite{apostol}
\begin{equation}
x_{s4}(t)=4a-2\sqrt{3a\varepsilon t}+2\varepsilon t
\end{equation}

The dependence on time of the corresponding shock velocity can now be found%
\begin{equation}
\frac{dx_{s4}}{dt}=2\varepsilon-\sqrt{\frac{3a\varepsilon}{t}}%
\end{equation}

\item \textbf{shock 5 ($x_{s5}(t)$)}, occurs between the two rarefaction
regions, at the time $t_{5}^{0},$ when the characteristic from the family
$x_{2}$ that originates at $x_{0}=0$ encounters $x_{s4}$
\begin{equation}
x_{r2}(\rho_{r1}=1/2-\varepsilon;t_{5}^{0})=x_{s4}(t_{5}^{0})
\end{equation}
resulting%
\begin{equation}
t_{5}^{0}=\frac{4a}{3\varepsilon}%
\end{equation}

The mass conservation conditions in this case ( see \ref{raref2} and
\ref{raref1}) reads
\begin{equation}
\frac{dx_{s5}}{dt}=\frac{j(\rho_{r2}(x_{s5}))-j(\rho_{r1}(x_{s5}))}{\rho
_{r2}(x_{s5})-\rho_{r1}(x_{s5})}=\frac{x_{s5}-2a}{t}\nonumber
\end{equation}
Considering that $x_{s4}(t_{5}^{0})=x_{s5}(t_{5}^{0})$ this can be integrated
to find
\begin{equation}
x_{s5}(t)=2a+(\varepsilon/2)t
\end{equation}
thus
\begin{equation}
\frac{dx_{s5}}{dt}=\varepsilon/2
\end{equation}

\item \textbf{shock 6 ($x_{s6}(t)$)}, occurs at $t_{6}^{0}$ when shock 5
encounters the characteristics from family $x_{5}$ that starts at $x_{0}=4a$.
Then, $t_{6}^{0}$ results from the relation$x_{s5}(t_{6}^{0})=4a.$ Using the
above derived expression for $x_{s5}(t)$ one gets
\begin{equation}
t_{6}^{0}=4a/\varepsilon
\end{equation}

The mass conservation conditions impose that
\begin{equation}
\frac{dx_{s6}}{dt}=\frac{j(\rho(x_{5}))-j(\rho_{r1}(x_{s5}))}{\rho(x_{5}%
)-\rho_{r1}(x_{s5})}=\frac{x_{s6}}{2t}\nonumber
\end{equation}
whose solution for the initial condition $x_{s6}(t_{6}^{0})=4a$ becomes
\begin{equation}
x_{s6}(t)=2\sqrt{a\varepsilon t}%
\end{equation}
from which one can find the shock velocity:
\begin{equation}
\frac{dx_{s6}}{dt}=\sqrt{\frac{a\varepsilon}{t}}%
\end{equation}

\end{itemize}

The fact that this velocity is always positive, implies that for
$\ t>t_{6}^{0}=4a/\varepsilon$ the solutions to Eq. (\ref{burgers 2}) for the
considered initial conditions (\ref{condinic}) are
\begin{align}
\rho(x,t)  &  =1/2\text{\hspace{0.15in}for \hspace{0.15in}}x<0\\
\rho(x,t)  &  =1/2-\frac{x}{2t}\text{\hspace{0.15in}for \hspace{0.15in}%
}0<x<4a+\sqrt{\frac{a\varepsilon}{t}}\\
\rho(x,t)  &  =1/2\text{ \hspace{0.15in}for}\hspace{0.15in}4a+\sqrt
{\frac{a\varepsilon}{t}}<x
\end{align}
From these, one sees that for long times, i.e. for $t>$ $t_{6}^{0}%
=4a/\varepsilon$, the density profile returns to its uniform steady state
profile, $\rho(x,t)=1/2.$This corresponds to the situation before perturbation
imposed by cargo particles. The solutions discussed above refer to behavior of
$\rho(x,t)$ within the transient regime soon after being perturbed by cargo.
The properties of shocks derived here are compiled in Table 1.

\newpage

Acknowledgments

\bigskip

DGL acknowledges the fellowship by the Brazilian agency Conselho Nacional de
Desenvolvimento Cient\'{\i}fico e Tecnol\'{o}gico (CNPq). CG acknowledges the
financial support from Funda\c{c}\~{a}o de Amparo \`{a} Pesquisa do Estado de
S\~{a}o Paulo (FAPESP).

\newpage{\LARGE Table 1}%
\[%
\begin{tabular}
[c]{|c|c|c|c|c|c|}\hline
\textbf{{shock number}} & \textbf{{initial time}} & \textbf{{time interval}} &
\textbf{{ travel distance}} & \textbf{average {velocity}} & \textbf{instant
{velocity}}\\\hline
$1$ & $0$ & $a/2\varepsilon$ & $a/2$ & $\varepsilon$ & $\varepsilon$\\
$2$ & $0$ & $a/2\varepsilon$ & $a/2$ & $-\varepsilon$ & $-\varepsilon$\\
$3$ & $a/2\varepsilon$ & $a/4\varepsilon$ & $0$ & $0$ & $0$\\
$4$ & $3a/4\varepsilon$ & $7a/12\varepsilon$ & $a/6$ & $2\varepsilon/7$ &
$2\varepsilon-\sqrt{\frac{3a\varepsilon}{t}}$\\
$5$ & $4a/3\varepsilon$ & $8a/3\varepsilon$ & $4a/3$ & $\varepsilon/2$ &
$\varepsilon/2$\\
$6$ & $4a/\varepsilon$ & $\infty$ & $\infty$ & \_ & $\sqrt{\frac{a\varepsilon
}{t}}$\\\hline
\end{tabular}
\ \ \ \ \ \ \
\]

\vspace{0.65in}

Table 1 - Kinematic properties of shock fronts formed within the course of
time evolution of the initial density profile in (\ref{condinic}). The average
velocity is calculated from the ratio between traveled distance (column 4) and
the corresponding time interval (column 3) for each processive movement.

\newpage

\medskip

\section{Figure Caption}

\bigskip

\textbf{Figure 1 -} The asymmetrical potential field introduced in Refs. (
\cite{astumian 96}, \cite{adjari prost}, \cite{magnasco}).

This is defined by parameters $a$ and $b$ and we chose $a>b.$

\vspace{0.2in}

\textbf{Figure 2 -} The motor density profile at different instants of time.
The cargo is represented by the circle and at each instant, its position is
adjacent to the shock front, the arrows pointing to the direction of the
movement. Fig.2(a) shows that the considered initial condition (\ref{condinic}%
) corresponds to a local perturbation on motor density with respect to its
steady state $\rho(x,t)=1/2.$

\textbf{Figure 3 -} Characteristics to Eq.(\ref{burgers 2}) for the initial
conditions given by (\ref{condinic}).

\end{document}